\documentclass[amsmath,amssymb,twocolumn]{revtex4}
\usepackage[parfill]{parskip}    
\usepackage{graphicx}
\usepackage{amssymb}
\usepackage{epstopdf}
\usepackage{color}
\usepackage{graphicx}
\usepackage{pdfpages}

\begin{document}

\title{Minimal Geometric Deformation decoupling in $2+1$ dimensional space--times}
\author{Ernesto Contreras {${}^{a}$\footnote{On leave 
from Universidad Central de Venezuela}
\footnote{ernesto.contreras@ciens.ucv.ve}}   
Pedro Bargue\~no{${}^{b}$}\footnote{p.bargueno@uniandes.edu.co}}
\address{
${}^a$Escuela Superior Polit\'ecnica del Litoral, ESPOL, Facultad de Ciencias Naturales y 
Matem\'atica, Campus Gustavo Galindo Km 30.5 V\'ia Perimetral, Guayaquil, Ecuador.\\
${}^b$Departamento de F\'{\i}sica,
Universidad de los Andes, Apartado A\'ereo {\it 4976}, Bogot\'a, Distrito Capital, Colombia
}
\begin{abstract}
We study the Minimal Geometric Deformation decoupling in $2+1$ dimensional space--times
and implement it as a tool for obtaining anisotropic solutions from isotropic geometries. 
Interestingly, both the isotropic and the anisotropic sector fulfill Einstein field equations 
in contrast to the cases studied in $3+1$ dimensions. In 
particular, new anisotropic solutions are obtained from the well known static BTZ solution.
\end{abstract}

\maketitle

\section{Introduction}\label{intro}

It is well known that the Minimal Geometric Deformation (MGD) decoupling, originally proposed 
\cite{ovalle2008} in the context of the Randall--Sundrum brane--world 
\cite{randall1999a, randall1999b}, has been a powerful tool to 
investigate self--gravitating distributions in the brane--world scenario
\cite{ovalle2009,ovalle2010,casadio2012,ovalle2013,ovalle2013a,casadio2014}
as well as to find new black hole solutions in a more general context 
\cite{casadio2015,ovalle2016} (for some recent appplications see for instance 
\cite{ovalle2015,casadio2015b,cavalcanti2016,casadio2016a,ovalle2017,
rocha2017a,rocha2017b,casadio2017a,ovalle2018,estrada2018,ovalle2018a,lasheras2018,gabbanelli2018,
sharif2018,fernandez2018,fernandez2018b}
).
In recent years, the use of the MGD--decoupling as a method to obtain new and relevant solutions 
of the Einstein field equations has 
increased considerably \cite{ovalle2017,ovalle2018,estrada2018,ovalle2018a,lasheras2018,gabbanelli2018,
sharif2018}. In particular, 
it is interesting to note that local anisotropy can be induced in well known spherically symmetric isotropic 
solutions of self--gravitating objects, leading to more realistic interior solutions of stellar systems.

Inspired by the success of the method in $3+1$ dimensional space--times,
it would be worth considering the application of the MGD- decoupling method in the lowest dimension in which the Einstein 
theory makes sense, {\it i. e.}, three dimensional space--times. Although, as stated by
Staruszkiewicz in his pioneering paper \cite{staruszkiewicz1963}
``three-dimensional gravitation theory is a theory without a field of gravitation; where no matter is present, space is flat'', 
solutions of the Einstein field equations in $2+1$ dimensional
space--times coupled to matter content have been considered as a testing ground to study
some aspects shared with their $3+1$ dimensional counterparts
with emphasis given to point particle solutions, perfect fluids, cosmological spacetimes, dilatons, inflatons,
stringy solutions, etc. (for a recent and very exhaustive review on 2+1 exact solutions, see \cite{Garciabook2017}).

In particular, some properties of
$3+1$-dimensional black holes such as horizons, Hawking radiation and black hole thermodynamics, are also present in
three-dimensional gravity which is simpler to deal with. Such is the
case of the celebrated BTZ \cite{btz} black hole solution, which shares many of the features
of the Kerr black hole, for instance the presence of event and inner horizons, an
ergosphere and a nonvanishing Hawking temperature.
\\
For these reasons, in this work we shall study the MGD-decoupling method in three-dimensional
space--times and obtain anisotropic solutions from the static BTZ solution. The
work is organized as follows. In the next section we review the main features
of the Einstein equations coupled to matter sources in three--dimensional 
space--times. Next, we implement the MGD-decoupling method applied to a circularly symmetric system containing a 
perfect fluid in section \ref{mgd}. Section \ref{BTZ} is devoted to obtaining anisotropic solutions from the static BTZ geometry. 
We summarize our conclusions in section \ref{remarks}.

\section{Einstein equations}\label{MGD}
Let us consider the Einsteins field equations
\begin{eqnarray}
R_{\mu\nu}-\frac{1}{2}R g_{\mu\nu}=-\kappa^{2}T_{\mu\nu}^{tot},
\end{eqnarray}
and assume that the total energy-momentum tensor is given by
\begin{eqnarray}
T_{\mu\nu}^{(tot)}=T_{\mu\nu}^{(m)}+\alpha\theta_{\mu\nu},
\end{eqnarray}
where $T^{\mu}_{\nu}=diag(-\rho,p,p)$ is the matter energy momentum for a perfect fluid and $\theta_{\mu\nu}$ is an additional source coupled with the perfect fluid by the constant $\alpha$.
Since the Einstein tensor is divergence free, the total energy momentum tensor $T_{\mu\nu}^{(tot)}$
satisfies
\begin{eqnarray}\label{cons}
\nabla_{\mu}T^{(tot)\mu\nu}=0.
\end{eqnarray}
In Schwarzschild like-coordinates, the circularly symmetric line element reads
\begin{eqnarray}\label{le}
ds^{2}=e^{\nu}dt^{2}-e^{\lambda}dr^{2}-r^{2}d\phi^{2},
\end{eqnarray}
where $\nu$ and $\lambda$ are functions of the radial coordinate $r$ only. 
Considering the metric (\ref{le}) as a solution of the Einstein equations, we obtain
\footnote{In what follows we shall assume $\kappa^{2}=8\pi$} 
\begin{eqnarray}\label{einst}
8\pi \tilde{\rho}&=&\frac{e^{-\lambda } \lambda '}{2r}\label{ein1}\\
8\pi \tilde{p}_{r}&=&\frac{e^{-\lambda } \nu '}{2r}\label{ein2}\\
8\pi \tilde{p}_{\perp}&=&-\frac{e^{-\lambda }}{4} \left(\nu ' \left(\lambda '-\nu '\right)
-2 \nu ''\right)\label{ein3},
\end{eqnarray}
where the prime denotes derivation respect to the radial coordinate and we have defined
\begin{eqnarray}
\tilde{\rho}&=&\rho+\alpha\theta^{0}_{0}\label{rot}\\
\tilde{p}_{r}&=&p-\alpha\theta^{1}_{1}\label{prt}\\
\tilde{p}_{\perp}&=&p-\alpha\theta^{2}_{2}\label{ppt}.
\end{eqnarray}
The conservation equation (\ref{cons}) reads
\begin{equation}\label{cons1}
p'+\frac{\nu'}{2}(\rho + p)-\alpha(\theta^{1}_{1})'+\frac{\nu'\alpha}{2}(\theta^{0}_{0}-\theta^{1}_{1})
+\frac{\alpha}{r}(\theta^{2}_{2}-\theta^{1}_{1})=0,
\end{equation}
which is a linear combination of Eqs. (\ref{ein1}), (\ref{ein2}) and (\ref{ein3}). Note that
Eqs. (\ref{ein1}), (\ref{ein2}) and (\ref{ein3}) correspond to the Einstein field 
equations for an anisotropic fluid. In this sense, the source $\theta_{\mu\nu}$ generate 
anisotropy in the original system controlled by the parameter $\alpha$, which 
disappears when $\alpha\to0$, as can be easily checked. 
Note that we have to solve for
Eqs. (\ref{ein1}), (\ref{ein2}), (\ref{ein3}) and (\ref{cons1}) but we deal with five unknows functions, 
$\{\nu,\lambda,\tilde{\rho},\tilde{p}_{r},\tilde{p}_{\perp}\}$. 
A conventional way to decrease the degrees of freedom to solve the system of differential equations considered
is providing an ansatz which in general is an equation of state relating
the components of the energy--momentum tensor. However, in this work, we shall obtain
solutions by the MGD--decoupling method, as explained further below.
\section{Minimal geometric deformation}\label{mgd}
In this section we introduce the MGD-decoupling method for $2+1$ dimensional space--times. 
Let us implement the following ``geometric deformation'' on the radial metric component
$g^{rr}$
\begin{equation}\label{def}
e^{-\lambda}=\mu(r)+\alpha f(r),
\end{equation}
where $\alpha$ is the decoupling parameter and $f(r)$ is the generic deformation undergone
by the radial metric component, $\mu(r)$. After replacing (\ref{def}) in Einstein equations
(\ref{ein1}), (\ref{ein2}) and (\ref{ein3}), we can separate the system 
of equations in two sets as follows. 
One set is obtained by setting $\alpha=0$ and corresponds to a perfect fluid 
\begin{eqnarray}
16\pi r \rho&=&-\mu '\label{iso1}\\
16\pi p&=&\frac{\mu \nu '}{r}\label{iso2}\\
32\pi p&=&\mu ' \nu '+2 \mu \nu ''+\mu \nu '^2\label{iso3},
\end{eqnarray}
with conservation equation given by
\begin{equation}\label{cons22}
p'+\frac{\nu'}{2}(\rho+p)=0,
\end{equation}
which is a linear combination of Eqs. (\ref{iso1}), (\ref{iso2}) and (\ref{iso3}). The other
set of equations corresponds to the source $\theta_{\mu\nu}$
\begin{eqnarray}
16\pi r \theta^{0}_{0}&=&-f'\label{aniso1}\\
-16\pi \theta^{1}_{1} &=&\frac{f \nu '}{r}\label{aniso2}\\
-32\pi \theta^{2}_{2}&=&f' \nu '+2 f\nu ''+f\nu '^2\label{aniso3},
\end{eqnarray}
with conservation
\begin{equation}\label{cons3}
(\theta^{1}_{1})'-\frac{\nu'}{2}(\theta^{0}_{0}-\theta^{1}_{1})-\frac{1}{r}(\theta^{2}_{2}-\theta^{1}_{1})=0.
\end{equation}
As in the previous case, Eq. (\ref{cons3}) is the linear combination of Eqs. (\ref{aniso1}),
(\ref{aniso2}) and (\ref{aniso3}). Note that
unlike the $3+1$ dimensional cases studied in 
\cite{ovalle2017,ovalle2018,ovalle2018a,lasheras2018,estrada2018} both the equations of the isotropic
(perfect fluid) and anisotropic ($\theta_{\mu\nu}$) sector are Einstein equations. In this
sense, the Einstein tensor $G_{\mu\nu}$ of the new solution turns out to be the linear combination of two Einstein tensor each one fulfilling Einstein field equations. More precisely, if
$G^{(m)}_{\mu\nu}=-\kappa^{2}T^{m}_{\mu\nu}$ stands for the isotropic sector, and 
$\tilde{G}_{\mu\nu}=-\kappa^{2}\alpha\theta_{\mu\nu}$ for the anisotropic one, the Einstein tensor
of the new solution is simply given by 
$G_{\mu\nu}=G^{(m)}_{\mu\nu}+
\tilde{G}_{\mu\nu}$. 
The above result can be naturally extended for arbitrary number of sources,
namely, given the Einstein Field Equations for a
collection of sources they can be transformed into a collection of Einstein's equations, one for each source. Even more, given a source $T^{(tot)}_{\mu\nu}=T_{\mu\nu}^{(m)}+\sum\limits_{i}\alpha_{i}\theta^{(i)}_{\mu\nu}$, 
with $i\ge 1$ and $\nabla_{\mu}T^{\mu\nu}=\nabla_{\mu}\theta^{(1)\mu\nu}=\cdots=\nabla_{\mu}\theta^{(n)\mu\nu}=0$, the Einstein tensor associated with $T^{(tot)}_{\mu\nu}$ can be decomposed as 
$G_{\mu\nu}=G^{(m)}_{\mu\nu}+G^{(1)}_{\mu\nu}+\cdots G^{(n)}_{\mu\nu}$
from where 
\begin{eqnarray}
G^{(m)}_{\mu\nu}&=&-\kappa^{2}T^{(m)}_{\mu\nu}\nonumber\\
G^{(1)}_{\mu\nu}&=&-\kappa^{2}\alpha_{1}\theta^{(1)}_{\mu\nu}\nonumber\\
\vdots & &\vdots\nonumber\\
G^{(n)}_{\mu\nu}&=&-\kappa^{2}\alpha_{n}\theta^{(n)}_{\mu\nu}.
\end{eqnarray}

This fact is remarkable because in $3+1$ space--times the anisotropic system does not fulfil Einstein
but ``quasi Einstein" field equations \cite{ovalle2017,ovalle2018,ovalle2018a} 
as a consequence of a missed $-\frac{1}{r^{2}}$ term which
avoid the matching with standard Einstein equations. Even more, in $3+1$ dimensions it is shown
that, despite this ``quasi Einstein" behaviour for the equations, the conservation can be written as a linear 
combination of the ``quasi Einstein" field equations and, therefore, the perfect fluid and the 
decoupling source $\theta_{\mu\nu}$ do not exchange energy but their interaction is purely gravitational, which can be summarized by
\begin{eqnarray}
\nabla_{\mu}{T}^{m\mu\nu}=\nabla_{\mu}\theta^{\mu\nu}=0.
\end{eqnarray}

In the next section we implement the MGD-decoupling method to obtain a new solution from the static
BTZ geometry.

\section{Anisotropic solution from the static BTZ geometry}\label{BTZ}
The static BTZ solution has a line element given by
\begin{eqnarray}\label{lel}
ds^{2}=(-M+\frac{r^{2}}{L^{2}})dt^{2}-\frac{dr^{2}}{-M+\frac{r^{2}}{L^{2}}}-r^{2}d\phi^{2},
\end{eqnarray}
from where
\begin{eqnarray}
e^{\nu}=\mu=(-M+\frac{r^{2}}{L^{2}}).
\end{eqnarray}
The matter content generating the static BTZ geometry is given by
\begin{eqnarray}
\rho&=& -\frac{1}{8\pi L^{2}}\\
p&=& \frac{1}{8\pi L^{2}}.
\end{eqnarray}

\begin{figure}[ht!]
\centering
\includegraphics[scale=0.7]{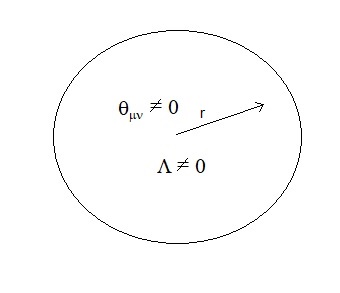}
\caption{\label{fig} 
Circularly symmetric space--time filled with both $\theta_{\mu\nu}$ and cosmological constant 
$\Lambda$. Note that the case $\theta_{\mu\nu}=0$ yields BTZ black hole.
}
\end{figure}

As it is well known, the static BTZ solution corresponds to a black hole 
in a space-time filled with a cosmological constant. In the next section we shall 
deform the BTZ black hole solution by
the MGD-decoupling method. More precisely, we shall fill the 
space--time with certain source $\theta_{\mu\nu}$ satisfying suitable equations of state \cite{ovalle2018a} 
which, after gravitational interaction with the cosmological constant, lead to the deformed BTZ 
geometry. In figure \ref{fig} we show schematically the kind of system we shall consider 
 henceforth.


\subsection{Isotropic solutions}
Considering an isotropic pressure for the source $\theta_{\mu\nu}$ \cite{ovalle2018a} implies
\begin{equation}
\theta^{1}_{1}=\theta^{2}_{2}.
\end{equation}
Combining Eqs. (\ref{aniso2}) and (\ref{aniso3}) leads to
\begin{equation}
f'+\frac{2 r f}{L^2 M-r^2}=0,
\end{equation}
from where
\begin{equation}
f=c_{1}L^{2}\left(-M+\frac{r^{2}}{L^{2}}\right),
\end{equation}
whith $c_{1}$ a constant of integration with dimension of inverse of length squared.
From Eq. (\ref{def}) we obtain 
\begin{eqnarray}\label{lambda1}
e^{-\lambda}=\left(1+\alpha c_{1}L^{2}\right)\left(-M+\frac{r^2}{L^2}\right).
\end{eqnarray}
Replacing in Eqs. (\ref{le}), (\ref{ein1}), (\ref{ein2})
and (\ref{ein3}) we obtain the line element
\begin{equation}\label{eli}
ds^{2}=(-M+\frac{r^{2}}{L^{2}})dt^{2}-\frac{dr^{2}}{
(\alpha c_{1}L^{2}+1) \left(-M+\frac{r^2}{L^{2}}\right)}-r^{2}d\phi,
\end{equation}
and the matter content
\begin{eqnarray}\label{mci}
\tilde{\rho}&=&-\frac{1+\alpha  c_{1} L^2}{8 \pi  L^2}\nonumber\\
\tilde{p}_{r}=\tilde{p}_{\perp}&=&\frac{1+\alpha  c_{1} L^2}{8 \pi  L^2}.
\end{eqnarray}

It is worth mentioning that Eqs. (\ref{eli}) and (\ref{mci}) correspond
to an isotropic solution for $2+1$ dimension. In fact, a straightforward calculation reveals that the curvature scalars read
\begin{eqnarray}\label{scalars}
R&=&6 \left(\alpha c_{1}+\frac{1}{L^2}\right)\nonumber\\
Ricc^{2}&=&\frac{12 \left(\alpha  c_{1} L^2+1\right)^2}{L^4}\nonumber\\
\mathcal{K}&=&\frac{12 \left(\alpha  c_{1} L^2+1\right)^2}{L^4},
\end{eqnarray}
where $R$, $Ricc^{2}$ and $\mathcal{K}$ stand form he Ricci, Ricci squared and the Kretschmann scalars
respectively. Note that from Eqs. (\ref{scalars}) we recover BTZ only if $\alpha\to0$. 

\subsection{Conformally symmetric solutions}
In this case we impose
\begin{equation}\label{conformal}
\theta^{2}_{2}=-\theta^{0}_{0}-\theta^{1}_{1},
\end{equation}
which implies that the source is traceless as required by 
the conformal symmetry. Combining condition (\ref{conformal}) with
(\ref{aniso1}), (\ref{aniso2}) and (\ref{aniso3}) we obtain
\begin{equation}
\frac{f'\left(L^2 M-2 r^2\right)}{r}+\frac{2 f\left(r^2-2 L^2 M\right)}{L^2 M-r^2}=0,
\end{equation}
from where
\begin{eqnarray}\label{sol2}
f=c_1\frac{ \left(r^2-L^2 M\right)}{\left(2 r^2-L^2 M\right)^{3/2}}.
\end{eqnarray}
In this case $c_{1}$ is a constant of integration with dimension of length.
In order to obtain the line element and the matter content of the 
new solution we perform the same procedure followed in the last section. In this
case combining (\ref{sol2}) and (\ref{def}) with (\ref{le}), (\ref{ein1}), (\ref{ein2})
and (\ref{ein3}), we obtain the $g^{rr}$ component of the metric
\begin{eqnarray}\label{lam}
e^{-\lambda}=\frac{\left(\frac{r^2}{L^2}-M\right) \left(\alpha  c_1 L^2+\left(2 r^2-L^2 M\right)^{3/2}\right)}{\left(2 r^2-L^2 M\right)^{3/2}},
\end{eqnarray}
and the anisotropic matter content given by
\begin{eqnarray}
\tilde{\rho}&=&\frac{L^2 r^2}{8 \pi  \left(L^3 M-2 L r^2\right)^2} \left(\frac{\alpha  c_1}{\sqrt{2 r^2-L^2 M}}4 M\right)\nonumber\\
&&-\frac{L^4 M}{8 \pi  \left(L^3 M-2 L r^2\right)^2}\left(\frac{2 \alpha  c_1}{\sqrt{2 r^2-L^2 M}}+M\right)\nonumber\\
&&-\frac{4 r^4}{8 \pi  \left(L^3 M-2 L r^2\right)^2}\\
\tilde{p}_{r}&=&\frac{1}{8 \pi }\left(\frac{\alpha  c_1}{\left(2 r^2-L^2 M\right)^{3/2}}+\frac{1}{L^2}\right)\\
\tilde{p}_{\perp}&=&-\frac{L^2 r^2}{8 \pi  \left(L^3 M-2 L r^2\right)^2} \left(\frac{\alpha  c_1}{\sqrt{2 r^2-L^2 M}}+4 M\right)\nonumber\\
&&+\frac{L^4 M}{8 \pi  \left(L^3 M-2 L r^2\right)^2} \left(M-\frac{\alpha  c_1}{\sqrt{2 r^2-L^2 M}}\right)\nonumber\\
&&+\frac{4 r^4}{8 \pi  \left(L^3 M-2 L r^2\right)^2}.
\end{eqnarray}
Note that in this case the method leads to an anisotropic solutions with 
curvature scalars given by
\begin{eqnarray}
R&=&\frac{6}{L^{2}}\\
Ricc^{2}&=&\frac{12}{L^4}-\frac{6 \alpha ^2 c_1^2 \left(L^4 M^2-L^2 M r^2+r^4\right)}{\left(L^2 M-2 r^2\right)^5}\\
\mathcal{K}&=&\frac{12}{L^4}-\frac{24 \alpha ^2 c_1^2 \left(L^4 M^2-L^2 M r^2+r^4\right)}{\left(L^2 M-2 r^2\right)^5}.
\end{eqnarray}
As in the previous case the BTZ solution is recovered
in the limit $\alpha\to0$. 

Now let us explore the causal structure of the solution. First note that the solution still have a Killing horizon ($e^{\nu} =0$) at $r_{H}=L\sqrt{M}$. Even more, 
$e^{-\lambda}$ diverges for the critical radius
\begin{equation}
r_{c}=L\sqrt{\frac{M}{2}}<r_{H}.
\end{equation}
In fact, this critical radius must be considered a real singularity provided some of 
the curvature scalar diverge at the same point. For this geometry we have two causal horizons at $e^{-\lambda} =0$.
The first one is at $r_{H}$, as in the BTZ case, but there is a second root 
given by
\begin{eqnarray}
r_{0}=\frac{\sqrt{\alpha ^{2/3} c_1^{2/3} L^{4/3}+L^2 M}}{\sqrt{2}}.
\end{eqnarray}
It is worth noticing that $r_{0}\ge r_{c}$ and depending on the values of the constant
will be inside or outside the Killing horizon $r_{H}$. In particular for
\begin{equation}\label{cond}
\alpha c_{1}<L M^{2/3},
\end{equation} 
we obtain that $r_{0}$ is in the interval $r_{c}<r_{0}<r_{H}$ and $r_{0}>r_{H}$ for $\alpha c_{1}>L M^{2/3}$. 
In order to avoid unacceptable space--time signature outside $r_{H}$ we should demand the
condition given by Eq. (\ref{cond}). In this case the behaviour resembles that of a charged 
BH solution but with singularity at certain critical radius, $r_{c}$, 
instead of at the origin. 

\subsection{Linear anisotropic source}
In this case, we choose a linear anisotropic source with equation of state
give by
\begin{equation}\label{linealsource}
\theta^{0}_{0}=a\theta^{1}_{1}+b\theta^{2}_{2}.
\end{equation}
Combination of (\ref{linealsource}) with
(\ref{aniso1}), (\ref{aniso2}) and (\ref{aniso3})  leads to 
\begin{equation}
\frac{2 f\left(L^2 M (a+b)-a r^2\right)}{L^2 M-r^2}+\frac{f' \left((b-1) r^2+L^2 M\right)}{r}.
\end{equation}
from where
\begin{equation}\label{sol3}
f=c_1 \left(r^2-L^2 M\right) \left((b-1) r^2+L^2 M\right)^{-\frac{a}{b-1}-1}.
\end{equation}
The constant of integration has dimensions 
of\linebreak
 $\textnormal{(Lenght)}^{-(2+\frac{2a}{1-b}-2)}$.
Combining (\ref{sol3}) and (\ref{def}) we obtain
\begin{equation}
e^{-\lambda}=F_{0}+\alpha  c_1 L^{4} \left(\frac{r^2}{L^{2}}- M\right) \left(\frac{(b-1) r^2}{L^{2}}+M\right)^{\Gamma}.
\end{equation}
where
\begin{equation}
F_{0}=\frac{r^2}{L^2}-M,
\end{equation}
and $\Gamma=-\frac{a}{b-1}-1$. To complete the analysis, we compute the 
content of the isotropic matter responsible of this 
geometry by replacing (\ref{sol3}) and (\ref{def})
in  (\ref{ein1}), (\ref{ein2}) and (\ref{ein3}) to obtain
\begin{eqnarray}\label{tm}
\tilde{\rho}&=&-\frac{1}{8 \pi  L^2}
-\frac{\alpha  c_1 \left(L^2 M (a+b)-a r^2\right)}{8 \pi}\times\nonumber\\
&&\times\frac{\left((b-1) r^2+L^2 M\right)^{\Theta}}{8 \pi }\\
\tilde{p}_{r}&=&
\frac{\alpha  c_1 L^2 \left(\frac{r^2}{L^2}-M\right)\left((b-1) r^2+L^2 M\right)^{\Gamma}}{8 \pi  L^2}\nonumber\\
&&+\frac{\left(\frac{r^2}{L^2}-M\right)}{8 \pi  L^2}\\
\tilde{p}_{\perp}&=&
-\frac{\alpha  c_1 L^2 F(z) \left((b-1) r^2+L^2 M\right)^{\Theta}}{8 \pi  L^6}\nonumber\\
&&+\frac{L^4 M+L^2 M r^2-2 L^2 r^2-r^4}{8 \pi  L^6},
\end{eqnarray}
where $\Theta=-\frac{a}{b-1}-2$ and
\begin{eqnarray}
F(z)&=&L^2 r^4 (a+(b-2) M-b+1)-(b-1) r^6\nonumber\\
&&+L^6 M^2+L^4 M r^2 (-a+M-2).
\end{eqnarray}
Now we analyse the causal structure directly from equations (\ref{tm}) \footnote{In this case the curvature scalars are too long expressions to be included in the manuscript.}.  First, note that we can avoid the apparition of any singularity taking $\Theta>0$ which implies
$b<1$ and $a>2(1-b)$.
Second, observe that $e^{-\lambda}$ has two roots: one corresponding to the Killing
horizon $r_{H}=L\sqrt{M}$ and the other at
\begin{equation}
r_{0}=\frac{\sqrt{\left(-\frac{1}{\alpha  c_1 L^2}\right){}^{\frac{a}{a+b-1}-1}-L^2 M}}{\sqrt{b-1}}.
\end{equation}
In this case, $r_{0}$ could be avoided for suitable choices
of the parameters $\alpha$, $c_{1}$, $a$ and $b$. In particular, note that no
real solutions can be obtained for $r_{0}$  when 
\begin{equation}
\left(-\frac{1}{\alpha  c_1 L^2}\right){}^{\Gamma}<L^2 M.
\end{equation}

\section{Conclusions}\label{remarks}
In this work we implemented the Minimal Geometric Deformation--decoupling method in
$2+1$ circularly symmetric and static space--times obtaining that both the isotropic and the anisotropic sector
fulfil Einstein field equations in contrast to the cases studied in $3+1$ dimensions, where
the anisotropic sector satisfies certain ``quasi--Einstein" field equations. In this sense the Einstein Field Equations for a
collection of sources can be transformed into a collection of Einstein's equations, one for each source.
As an example, we implemented the decoupling method to obtain new solutions from
the well known static BTZ geometry. In particular, the anisotropic system were solved
providing suitable equations of state for the source $\theta_{\mu\nu}$ namely the isotropic, the conformal 
and the linear equation of state. The results are in concordance with their $3+1$ counterparts obtained in reference
\cite{ovalle2018a} in the sense that some extra structures such as causal horizons and singularities
appear as a consequence of the Minimal Geometry Deformation-decoupling. In adition, it was shown that in the
case of linear equation of state those extra structures can be avoided for certain values of the free
parameters of the solution, as shown in reference \cite{ovalle2018} for 3+1 spacetimes.\\ 
We conclude this paper by noting that the method here developed can be easily 
applied to obtain new and relevant solutions taking as the isotropic sector any of the 
already known 2+1 space--times.

\section*{ACKNOWLEDGEMENTS}
The authors would like to acknowledge Jorge Ovalle for fruitful discussions and correspondence. The author P.B. was supported by the Faculty of Science
and Vicerrector\'ia de Investigaciones of Universidad de
Los Andes, Bogot\'a, Colombia.

\end{document}